# Polarized neutron reflectometry study of $Fe_{16}N_2$ with Giant Saturation Magnetization prepared by N Inter-diffusion in Annealed Fe-N Thin Films

Nian Ji[1,2], Valeria Lauter[3], Hailemariam Ambaye[3], Xiaowei Zhang[1,2], Sarbeswar Sahoo[4], Mike Kautzky[4], Mark Kief[4] and Jian-Ping Wang[1,2,*]

[1]*The Center for Micromagnetics and Information Technologies (MINT) & Electrical and Computer Engineering Department, University of Minnesota, Minneapolis, Minnesota 55455*

[2]*School of Physics and Astronomy, University of Minnesota, Minneapolis, Minnesota 55455*

[3]*Neutron Science Scattering Division, Oak Ridge National Laboratory, Oak Ridge, Tennessee 37831*

[4]*Seagate Technology, Minneapolis, Minnesota, 55435*

## Abstract

We report a synthesis route to grow iron nitride thin films with giant saturation magnetization ($M_s$) through an N inter-diffusion process. By post annealing Fe/Fe-N structured films grown on GaAs(001) substrates, nitrogen diffuses from the over-doped amorphous-like Fe-N layer into strained crystalline Fe layer and facilitates the development of metastable $Fe_{16}N_2$ phase. As explored by polarized neutron reflectometry, the depth-dependent $M_s$ profile can be well described by a model with the presence of a giant $M_s$ up to 2360 emu/cm$^3$ at near-substrate interface, corresponding to the strained regions of these annealed films. This is much larger than the currently known limit ($Fe_{65}Co_{35}$ with $M_s$~ 1900 emu/cm$^3$). The present synthesis method can be used to develop writer materials for future magnetic recording application.

* email: jpwang@umn.edu

α"-$Fe_{16}N_2$ is one of the most enigmatic magnetic materials over the last several decades. The apparent conflict centers on its widely varying saturation magnetization ($M_s$) observed by different researchers using traditional magnetometry methods (e.g. vibrating sample magnetometry)[1, 2, 3, 4, 5, 6, 7, 8]. Recently, we have provided a first-principle calculation based on a proposed "($Fe_6N$) cluster + (Fe) atom" model[9], which shows an unconventional scenario to establish $Fe_{16}N_2$ with giant $M_s$. However, on one hand, there is still lack of direct convincing experimental evidence to confirm the existence of its giant $M_s$. On the other hand, special layer structures and advanced materials processing conditions are highly demanded for its potential application in magnetic recording writing head, given the relatively high magnetocrystalline anisotropy of this material.

In this paper, we used polarized neutron reflectometry (PNR) to *directly* explore the depth-dependent saturation magnetization profile of annealed samples with (Fe/Fe-N) and (Fe/Fe-N) $_2$ structures on GaAs substrate. It was found that a giant $M_s$ can be established at the near substrate or Fe/Fe-N interface. This is attributed to the ordered occupation of N in the interstitial sites upon N inter-diffusion and consequently, to the formation of $Fe_{16}N_2$ phase at these interfacial regions after a long time post annealing. The highest $M_s$ observed here (~ 2360 emu/cm³) is consistent with the pure phase $Fe_{16}N_2$ films prepared by molecule beam evaporation (MBE) method (Ref. 2). Unlike any previously reported synthesis processes of $Fe_{16}N_2$ films requiring either heated substrates (Ref. 2, 3) or involving phase-transition from the as-deposited $Fe_8N$ films into $Fe_{16}N_2$ (Ref. 4, 5, 6, 7, Error: Reference source not found), the synthesis approach described here utilizes an inter-diffusive nitrogenation that N was purposely over-doped in FeN layer, which was found close to be amorphous at as-deposited state. The subsequent annealing process allows the N atoms diffuse into its adjacent strained Fe crystalline layer to form $Fe_{16}N_2$ phase (Fig. 1a) and stabilize this chemically ordered phase.

The bi-layer Fe/Fe-N structured sample was prepared by first depositing a 2 nm Fe layer on GaAs (001) surface at substrate temperature of 250 °C, then cooling down to room temperature and adding another 12 nm Fe-N layer by sputtering Fe targets with thoroughly mixed Ar+$N_2$ with an excess of $N_2$ partial pressure[10]. The [Fe/Fe-N)$_2$/GaAs sample was similarly obtained by first depositing a 2 nm Fe on GaAs (001) at 250°C, then adding a sandwiched structure of Fe-N(12nm)/Fe(2nm)/Fe-N(12nm) after cooling down to room temperature. The x-ray diffraction (XRD) pattern of the as-deposited bi-layer sample (Fig 1b top panel) reveals a broad bump ranging from 45° ~ 52°. This can be attributed to the presence of N rich Fe-N phase and is nearly amorphous, which is similar to the previous report on amorphous Fe-N system[11]. After post-annealing the as-deposited films at substrate temperature of 120 °C for 40 hrs, the XRD data of the final products are shown in the bottom panel of Fig. 1b. Both the annealed bi-layer (labeled as S1) and annealed four-layer (labeled as S2) samples show a similar XRD pattern with exclusive *(00l)* orientation. The chemical ordering of N site is evident from the superlattice diffraction (002), which is consistently seen in all the previous reports (Ref. 2~8).

The low-angle grazing incident x-ray reflectivity (GIXR) curves are measured and shown in Fig. 1c for both samples (S1 and S2), which are subsequently analyzed to retrieve the depth structural information. The calculated reflectivity curves that best reproduce the experimental data are shown with their x-ray electron density (XED) depth profiles plotted in the inset. When fitting the reflectivity curves of both samples, a three layer model is used to reproduce the experiment data, indicating a not uniform chemical structural through the depth direction due to the distribution of nitrogen. The XED of the substrate is fixed at a value close to that of a nominal GaAs ($3.84 \times 10^{-2}$ nm$^{-3}$). The XED of the film expectedly reproduced a value of bulk Fe ($5.88 \times 10^{-2}$ nm$^{-3}$)[12], followed by a FeN layer with a slightly reduced XED.

PNR experiments were conducted on the Magnetism Reflectometer on beamline 4A at Spallation Neutron Source, Oak Ridge National Laboratory [13]. Fig. 2 shows the non-spin-flip (NSF) specular reflectivity as a function of wave vector transfer [Q= $(4\pi/\lambda)\sin\theta$, where $\theta$ is the incident angle on the film and $\lambda$ is the wave length in (nm) of the neutron beam measured on samples S1 and S2 (the exact same pieces as measured by GIXR) at room temperature in a saturation external magnetic field. In the PNR data, the two curves R+ and R- correspond to the reflected intensity of neutrons with the spins either parallel (R+) or anti-parallel (R-) to the direction of the external magnetic field ($\mu_0 H=1.0$ T) applied in plane of the sample. In general, to do PNR data fitting, the depth information from both magnetic and chemical structure are allowed to vary in order to reproduce the experimental data. To ensure the accurate determination of the depth profile calculation in our case, the structural information from the GIXR results (Fig. 1c) was used for the refinement of the PNR data. It is known that co-refining the x-ray and neutron reflectivity curves allows an unambiguous determination of magnetic depth profile[14, 15]. Therefore, in modeling the chemical part (nuclear scattering length density (NSLD)) of the PNR results, we set the initial structural information to be consistent with that of the GIXR results, allowing marginal adjustment during the fitting process. Following this analysis procedure, the only parameter being allowed for free evolution is the magnetic part. This finally gives minor but appreciable difference in NSLD between layers (black lines in Fig. 2c and e), which is expected due to the comparable scattering cross-section between N and Fe for neutron probe. In addition, this observation also suggests the N concentration at the layer interfacial region is lower than that in the initial Fe-N layers.
4

The key results are presented in the magnetic scattering length density (MSLD) depth profiles (red lines in Fig. 2d and f). There are several interesting features. In sample S1, the MSLD reaches a highest value of 6.76 x$10^{-6}$ Å$^{-2}$, corresponding to a $M_s$ of 2360 emu/cm$^3$ [16] at the bottom interface between the substrate and film within a thickness region of 50~70 Å, This value of magnetization is remarkable and substantially larger than that of $Fe_{65}Co_{35}$ (5.7 x$10^{-6}$ Å$^{-2}$). Towards the top interface, the magnetization abruptly reduces and reaches a plateau with $M_s$ of 1300 emu/cm$^3$. This value is much smaller than the bulk value of Fe of about ~ 1700 emu/cm$^3$ and cannot be solely attributed to the disordering effect of N in the stoichiometry identical $Fe_8N+Fe_{16}N_2$ system, in which case, the $M_s$ of $Fe_8N$ is in general at least larger than 1700 emu/cm$^3$. More likely, this reduction of $M_s$ is due to the presence of some N-rich Fe-N phase that exists near the top interface given the synthesis procedure and the observed enhancement of NSLD at corresponding regions. It is also appealing to notice that the magnetic interface between the high and low magnetization region is sharp, suggesting two well separated iron nitride regions along the direction of surface normal, in which we ascribe them as $Fe_{16}N_2$ and N-rich Fe-N, respectively. In sample S2, similar magnetic configuration is also shown. A reduced but still "giant" magnetization resides at the bottom interface (MSLD =6.35 x$10^{-6}$ Å$^{-2}$), corresponding to the $M_s$ of 2200 emu/cm$^3$. However, different from sample S1, the transition region is more gradual and at the interface between Fe and Fe-N towards the top, a broad peak feature is seen for the MSLD, suggesting an enhancement of magnetization at this region, accompanied with a slightly increased value of the plateau with $M_s$ = 1380 emu/cm$^3$ near the top interface. The $M_s$ value of the middle portion of sample S2 is between 1400 and 1700 emu/ cm$^3$. Since the structural analysis suggests that the only possible Fe-N phases are α''-$Fe_{16}N_2$ and α'-Fe-N martnesites for both samples, these observations can only be appreciated by considering the formation of mixture phases consisting of $Fe_{16}N_2$ and N-rich Fe-N in the region as highlighted, likely due to the complicated strain effect given the multiple interfaces in contrast to that in sample S1. These observations are consistent with our recent PNR experiments on MgO substrate based samples[17].

It is worth mentioning that a single layer model with both MSLD and NSLD to be uniform throughout the film fails to describe the essential features of the reflectivity curves measured of the present samples. As shown in Fig. 3, we compare PNR data of sample S1 using two different models to fit the experimental data (a) and (b). We also show the PNR result of a reference sample of a single Fe film grown on GaAs substrate using the same facing target sputtering method. Similar GIXR measurement was also conducted on the Fe sample to co-refine the structural depth profile of its PNR data. The corresponding depth-profiles are plotted in (e)~(f). In the uniform "normal $M_s$ model", we set the MSLD to be equal to that of nominal Fe (~5 x$10^{-6}$ Å$^{-2}$). It is clearly seen a large discrepancy between the experiment and calculation on the R$^-$ curve as well as the high Q region for both R+ and R- curves. This can be understood by looking at the obtained SLD for these two models and comparing with that of Fe reference sample. In the Fe-N sample, the fringe features of the R- curve are more pronounced in contrast to that in Fe sample. Since the magnitude of the oscillation is proportional to the contrast of SLD between film and substrate, these observations suggest that in the Fe film, for the R$^-$ curve, the SLD is similar for the film and substrate. However, this is not the case in the sample S1. As illustrated, a relatively uniform SLD with its value calculated to be close to that of GaAs substrate is seen from the R- SLD profile of the Fe sample. In contrary, a more complicated SLD structure is obtained in describing the R- reflectivity curve of the Fe-N sample.

To further estimate the limiting values of the magnetic depth profile obtained, especially for the high Ms region at the bottom interface, we compared the experimental PNR data with three model calculations shown in Fig 4 with dots and solid lines, respectively. In model A, the NSLD and MSLD profiles are the same as in Fig. 2 and 3, which shows a giant $M_s$. In model B and C, we reduced the MSLD of the bottom layer by 10% and 20% respectively, while keeping the structural profile, which was obtained from the X-rays and neutron experiment, fixed. It is seen as we progressively reduce the Ms of the bottom layer, the calculation becomes increasing worse to describe the experimental data as evidenced by the arrows shown in Fig. 4a. Furthermore, if allowing the NSLD for free evolution while fixing the MSLD profile in model B and C, the calculation quickly converges with only marginal modification on the NSLD as well as the reflectivity curves. Additionally, we also studied a model with MSLD constrained within $5.9 \times 10^{-4} nm^{-2}$ (this is slightly above the Ms of FeCo~1900 emu/cm$^3$). All other parameters, including thickness, roughness and NSLD are allowed to evolve freely. As it is shown in Fig 5a, it is possible to produce equally good fit to the experimental data. But the NSLD looks drastically different comparing to model A. The two layers have large NSLD contrast. Then, we used this NSLD profile to calculate the XED profile (assuming the NSLD is proportional to the mass density) and x-ray reflectivity curves as shown in (c) and (d). It is clearly seen that this density profile can not describe the experimental data.

In summary, we have demonstrated that Fe-N thin film prepared by the N inter-diffusive process can develop $Fe_{16}N_2$ phase. The PNR data obtained on these samples show a giant Ms at the part of the film at the bottom interface. From the practical viewpoint, the direct use of $Fe_{16}N_2$ as a magnetic recording write-head material could be limited by its relatively large $H_k$ given its tetragonal lattice structure[18]. The layer structure presented and discussed above allowed the formation of FeN layer with a reduced anisotropy while maintain a relative high $M_s$, which can function as the writing pole. This may provide an alternative to the current FeCo based write-head material. It is interesting to note that the achieved composite layer with graded $M_s$ also provides design flexibility for the Shingle magnetic writer[19].

This work was partially supported by Seagate Technology, including the sample preparation and XRD characterization. The PNR testing and analysis by NJ, VL and HA done at SNS, ORNL are sponsored by U.S. Department of Energy, Office of Basic Energy Sciences under No. DE-AC02-98CH10886. The authors would like to thank Prof. R. Victora, Prof. P. Crowell and Prof. C. Leighton for useful discussion.

Figure caption

FIG. 1 (a) A schematic illustration of the formation of $Fe_{16}N_2$ due to the N diffusive process. (b) XRD θ-2θ scans of sample S1 and S2 (c) Fitted GIXR data of sample S1 ad S2. (Data vertically shifted one order of magnitude for clarity). The inset shows depth-dependent x-ray electron density (XED) profile obtained from the fit to the experimental data

FIG. 2 Fitted spin-up (R+) and spin-down (R-) polarized neutron reflectivity curves for sample S1 (a) and S2 (b), respectively; Magnetic (bottom) and structural (upper) SLD for sample S1 (c) (d) and S2 (e) (f). The dashed lines represent the Magnetic SLD of $Fe_{65}Co_{35}$ (currently known as the material with highest saturation magnetization so far). The regions correspond to different phases are outlined and explained in text

FIG. 3 Fitted spin-up (R+) and spin-down (R-) polarized neutron reflectivity curves for (a) sample S1 using non-uniform high Ms model, (b) sample S1 using uniform normal Ms model, (c) Fe reference sample. (e)~(f) show the corresponding $R^+$ (red) and $R^-$ (blue) SLD as calculated from the models.

FIG. 4 (a) spin-up (R+) and spin-down (R-) polarized neutron reflectivity experimental data (grey dots) and model A (red), B (blue) and C (green) calculated curves. (b) The corresponding MSLD profile as shown in the model calculation and NSLD profile (brown). The dashed line shows the MSLD value of bulk $Fe_{65}Co_{35}$, known as the material with highest Ms so far.

Figure 5 (a) Experimental PNR data (grey dots) and model A, D calculations based on the depth profiles shown in the (b). (c) Experimental XRR data (grey dots) and calculations based on the model A and D with XED depth profile shown in (d).

FIG. 1

FIG. 2

FIG. 3

FIG .4

FIG. 5